\documentclass{article}
\usepackage{spconf,amsmath,graphicx}
\usepackage{balance}

\usepackage{amsmath}

\usepackage{amsthm}
\usepackage{amssymb}
\usepackage{amsfonts}
\usepackage{dsfont}
\usepackage{accents}
\usepackage{bm}
\usepackage{hyperref}

\usepackage{tabularx} 
\usepackage{caption}
\usepackage{multirow}
\usepackage{booktabs}
\usepackage{graphicx}
\graphicspath{{Figures/}}
\usepackage{subcaption}
\usepackage{tcolorbox}
\usepackage{wasysym}
\usepackage{longtable}
\usepackage{booktabs}

\usepackage{tikz,pgfplots}
\usepgfplotslibrary{groupplots}
\usetikzlibrary{intersections,calc,arrows,matrix,spy}
\usepackage{color}
\definecolor{darkred}{rgb}{0.6,0,0}
\definecolor{darkgreen}{rgb}{0,0.5,0}
\definecolor{darkblue}{rgb}{0,0,0.5}
\definecolor{SkyBlue}{rgb}{0.53, 0.81, 0.92}
\pgfplotsset{compat=1.5.1}

\usepackage{algorithm}
\usepackage{algpseudocode}
\usepackage[squaren,Gray]{SIunits}

\usepackage{placeins}
\usepackage{float}

\definecolor{mycolor1}{RGB}{47, 89, 245}%
\definecolor{mycolor2}{RGB}{214, 101, 41}%
\definecolor{mycolor3}{RGB}{19, 161, 11}%
\definecolor{mycolor4}{rgb}{0.,0.,1}%

\definecolor{sid}{RGB}{225, 0, 100}

\def\SNR{\mathrm{SNR}}
\def\R{\mathbb{R}}          									 	          

\newcommand{\xb}{\mathrm{\mathbf{x}}}
\newcommand{\yb}{\mathrm{\mathbf{y}}}

\newcommand{\Xb}{\mathrm{\mathbf{X}}}

\newcommand{\gammab}{\mathrm{\bm{\gamma}}}

\newcommand{\etab}{\mathrm{\bm{\eta}}}
\newcommand{\thetab}{\mathrm{\bm{\theta}}}

\newcommand{\rhob}{\mathrm{\bm{\rho}}}

\newcommand{\phib}{\mathrm{\bm{\phi}}}

\newcommand{\Phib}{\mathrm{\bm{\Phi}}}

\newcommand{\proj}{\mathbf{P}}
\newcommand{\rot}{\mathbf{R}}

\newcommand{\var}{\mathrm{Var}}
\newcommand{\D}{\mathbf{D}}

\newcommand{\Lc}{\mathcal{L}}

\usepackage{tcolorbox}
\usepackage{mdframed}
\usepackage{lipsum}

\newcommand{\eqdef}{\ensuremath{\stackrel{\mbox{\upshape\tiny def.}}{=}}}

\graphicspath{{./images/}}

\title{Joint Cryo-ET Alignment and Reconstruction with\\Neural Deformation Fields}

\name{
Valentin Debarnot$^\dagger$, 
Sidharth Gupta$^*$, 
Konik Kothari$^*$,
and
Ivan Dokmani\'c$^{\dagger\,*}$
\thanks{This research was supported by the European Research Council (ERC) Starting Grant 852821-SWING.}
\thanks{The authors wish to thanks Ben Engel, Ricardo Diogo Righetto and Lorenz Lamm for preliminary discussion about the cryo-ET problem.}}
\address{
$^\dagger$University of Basel, 
$^*$University of Illinois at Urbana-Champaign	
}

\begin{document}

\maketitle 
\begin{abstract}
	We propose a framework to jointly determine the deformation parameters and reconstruct the unknown volume in electron cryotomography (CryoET). CryoET aims to reconstruct three-dimensional biological samples from two-dimensional projections. A major challenge is that we can only acquire projections for a limited range of tilts, and that each projection undergoes an unknown deformation during acquisition. Not accounting for these deformations results in poor reconstruction. The existing CryoET software packages attempt to align the projections, often in a workflow which uses manual feedback. Our proposed method sidesteps this inconvenience by automatically computing a set of undeformed projections while simultaneously reconstructing the unknown volume. We achieve this by learning a continuous representation of the undeformed measurements and deformation parameters. We show that our approach enables the recovery of high-frequency details that are destroyed without accounting for deformations. 	\end{abstract}


\section{Introduction}

Tomographic imaging plays a central role in science, medicine, and engineering. An emerging representative in biological imaging is electron cryotomography (CryoET). Unlike single-particle cryoelectron microscopy, CryoET can image entire cells under cryogenic conditions. The three-dimensional volume is tilted around an axis relative to a probing electron beam. A sensor array then collects a series of two-dimensional projections---a \textit{tilt-series}---at a discrete set of tilt angles.

More formally, we measure $M$ projection images of resolution $N \times N$ 
of an unknown volume $\rhob \in \R^{N \times N \times N}$, 
\begin{align}
    \yb_m = \D(\phib_m^\star)\proj \rot(\theta_m) \rhob + \etab_m, \quad m = 1, \ldots, M . \label{eq:CryoET_formulation}
\end{align}
In \eqref{eq:CryoET_formulation} $\rot(\theta_m)$ denotes the rotation (tilt) by an angle $\theta_m$ and $\proj$ denotes a projection from $\R^{N\times N\times N}$ to $\R^{N\times N}$ which is a simple summation over the last coordinate. Due to the mechanical stage drift and beam-induced sample motion, the CryoET projections are affected by deformations such as shifts, shears, and rotations~\cite{mastronarde2007fiducial}. We model the deformations by the operator $\D(\phib_m^\star)$, with $\phib_m^\star$ the deformation parameters for the $m$th projection. The noise $\etab_m$ is iid Gaussian. We aim at recovering $\rhob$ from $\{\yb_m\}_{m=1}^M$. However, we face two challenges: 1) the tilt $\theta_m$ can only vary between -70 and +70 degrees resulting in a \textit{missing wedge} of measurements, and 2) the deformation parameters, $\phib_m^\star$, are unknown.

If we ignore the deformation operator, the reconstructed volume (\textit{tomogram}) can be obtained by filtered-back projection (FBP) \cite{harauz1986exact}. 
In order to account for the unknown deformations, many popular CryoET reconstruction packages first perform a tilt series alignment in order to invert the degradation caused by $\D(\phib_m^\star)$ \cite{mastronarde2017automated,fernandez2021tomoalign,tegunov2019real}. These packages show that deformation estimation is vital for accurate reconstruction, but they are often based on geometric heuristics which do not guarantee an optimal reconstruction. This motivates our work: we build a framework to jointly recover the deformation parameters and the unknown volume which minimize a data consistency loss. We adapt the recent framework of Gupta et al. \cite{gupta2022differentiable} to leverage neural fields (or implicit neural representations) \cite{xie2022neural} with their key property that they enable automatic differentiation with respect to input coordinates. The coordinate-based framework allows us to effectively parameterize various classes of deformations.

\subsection{Related work}

Existing CryoET software such as IMOD \cite{mastronarde2017automated,tegunov2019real}, TomoAlign \cite{fernandez2021tomoalign}, and Warp \cite{tegunov2019real} handles deformation using fiducial markers in the specimen. Some recent software such as AreTomo estimates the deformation parameters without
fiducial markers \cite{zheng2022aretomo}, by leveraging the geometry of structured misalignments and by tracking patches from tilt to tilt. While this leads to an automated reconstruction pipeline, it is still based on heuristics. 
What is more, we could not find a precise mathematical description of the method and there exists no open source code \footnote{At the time of writing, this seems to hold more generally: there are no open source CryoET packages that handle deformations.} 
A recent work by Liu et al. \cite{liu2022mechanical} that appeared during preparation of this manuscript similarly proposes to jointly estimate the unknown object and the deformation parameters in optical tomography. In this paper, we represent the deformation parameters and the projection images by a continuous neural field which allows us to optimize over them using standard optimizers (without alternating between the deformation and the volume), while being able to plug-and-play almost arbitrary deformations.

Neural fields (or implicit neural networks) represent continuous signals as maps from coordinates to function values \cite{xie2022neural}. Their use cases include fitting 3D radiance maps to 2D images \cite{mildenhall2020nerf} and solving partial differential equations \cite{sitzmann2020implicit}. Sun et al.  applied implicit networks to interpolate and upsample measurements in 2D computed tomograph \cite{sun2021coil}. Rather than to obtain denser measurements, we use automatic differentiation \textit{with respect to coordinates} to optimize over parameters in the measurement space and thus fit the \textit{unobserved} measurements. 


\section{Learning deformations}

We represent the measurements $\{\yb_m\}_{m=1}^M$ in \eqref{eq:CryoET_formulation} by an implicit neural network~\cite{mildenhall2020nerf}. These networks parameterize a continuous representation of the observed measurements. Automatic differentiation, available in all major deep learning libraries, then allows us to compute the gradients of this continuous representation with respect to the measurement coordinates---a three-dimensional coordinate comprising a tilt angle and a location on the two-dimensional sensor array. In this work, we train an implicit neural network, $f_\gammab:[-\pi,\pi)\times \R^2 \to \R$, parameterized by $\gammab\in\Gamma$, where $\Gamma$ is the space of feasible parameters. For each tilt angle $\theta_m$ we denote by $\Xb$ a uniform sampling of the sensor array.
We train the network to reproduce the undeformed measurements, that is, for each point $\xb\in\Xb$ of the sensor array grid we want
\begin{equation}
f_\gammab(\theta_m, \xb) \approx (\proj \rot(\theta_m) \rhob)(\xb).
\end{equation}
Importantly, since the non-deformed projections $\proj \rot(\theta_m) \rhob$ cannot be observed, we also learn the unknown deformation parameters $(\phib_m^\star)$ jointly with the implicit neural network. This is the crux of the proposed method: since the deformation parameters are functions of the measurements coordinates (tilt and position within a projection), we can directly use automatic differentiation to optimize over them, simultaneously with optimizing the weights $\bm{\gamma}$ of the neural field network. The objective we minimize has two components. The first one is simply the usual interpolation (fitting) loss subject to \textit{unknown} deformations,
\begin{equation}
\Lc_{\mathrm{data}}(\phib,\gammab) \eqdef \sum_{m=1}^M \| \D(\phib_m)f_\gammab(
\theta_m, \cdot) - \yb_m\|_{\ell^2(\Xb)}^2.
\end{equation}
Here
$\phib$ denotes all deformation parameters, $\phib_1, \ldots, \phib_M$, and
$\|\cdot\|^2_{\ell^2(\Xb)}$ is the usual $2$-norm computed with $f_\gammab$ sampled on $\Xb$. 

Minimizing $\Lc_{\mathrm{data}}(\phib,\gammab)$ alone is clearly insufficient to learn correct deformations. We thus restrict the class of implicit network with parameters $\gammab$ so that the deformation parameters can be estimated within a physically meaningful range,
\begin{equation}
\Lc_{\mathrm{op}}(\gammab) \eqdef \| f_\gammab -  A(\thetab)A^\dagger(\thetab)f_\gammab\|_{\ell^2( \Xb)}^2,
\end{equation}
where $A(\thetab)\eqdef [a(\theta_1),\hdots,a(\theta_M)]$ with $a(\theta_m) \eqdef \proj \rot(\theta_m)$ and  $A^\dagger(\thetab)$ denotes the filtered backprojection. 

This method is inspired by recent work on implicit representations for correcting operator error~\cite{gupta2022differentiable} and extends the framework to handle random deformations seen in practice in CryoET.
Since CryoET operates at very low SNRs we use an additional total variation norm regularizer,
\begin{equation}
\Lc_{\mathrm{reg}}(\gammab) \eqdef \lambda_\theta \|\nabla_{\theta} f_\gammab \|_{\ell_1( \Xb)} + \lambda_\xb\|\nabla_{\xb} f_\gammab \|_{\ell_1( \Xb)},
\end{equation}
where $\nabla_{\theta} f_\gammab$ and $\nabla_{\xb} f_\gammab$ refer respectively to the gradient of $\theta \mapsto f_\gammab(\theta,\cdot)$ and $\xb \mapsto f_\gammab(\cdot,\xb)$.
We empirically verify that this term helps obtain accurate reconstructions. The regularization of the implicit network along the coordinate modeling the tilt angle is important to correctly estimate the parameters $\phib$. It ensures that two estimates of consecutive undeformed observations do not differ drastically. 
Our experiments suggest that a small value of $\lambda_\xb$ and $\lambda_\theta$ suffices to stabilize the joint 
measurement representation and deformation parameter learning
(cf. Section~\ref{sec:numerics}). 

Summarizing, we compute $(\widehat\gammab, \widehat \phib)$ that solve
\begin{equation}
\label{eq:optim}
\min_{\gammab\in\Gamma, \phib\in \Phib}  \lambda_{1} \Lc_{\mathrm{data}}(\phib,\gammab) +  \lambda_{2}\Lc_{\mathrm{op}}(\gammab) + \Lc_{\mathrm{reg}}(\gammab),
\end{equation}
where $\Phib$ is the space of admissible deformations and $\lambda_1,\lambda_2\geq0$ and empirically chosen. 
We use the Adam algorithm to minimize \eqref{eq:optim}. The final tomogram is given by
\begin{equation}
\widehat{\rhob} \eqdef A^\dagger(\thetab) f_{\widehat{\gammab}}.
\end{equation}

One advantage of our approach is that we do not need to explicitly invert the deformation operator: a potentially problematic task for nonlinear or non-invertible deformations. Furthermore, our framework allows us flexibility in budgeting computation. For example, if repeatedly applying the forward operator to a high-resolution volume is computationally expensive, we can initially set $\lambda_2=0$ to obtain a coarse measurement representation and then refine it with $\lambda_2 > 0$.

\section{Experiments}\label{sec:numerics}


\subsection{Influence of noise on individual cells}

We use the volume density of a native M. pneumoniae cell treated with chloramphenicol \cite{tegunov2021multi} (dataset DOI on EMPIAR 10.6019/EMPIAR-10499); see Fig. \ref{fig:molecule_volume}.

\subsubsection{Experimental parameters} \label{sec:exp_params_molecule}

We simulate CryoET acquisition using $60$ projections at angles between $-70$ and $+70$ degrees and using the deformation and noise model \eqref{eq:CryoET_formulation} at various signal-to-noise ratios (SNRs),
\begin{equation*}
\SNR(\yb_0, \etab) = 10\log_{10}\left(\frac{\var(\yb_0)}{\var(\etab)}\right).
\end{equation*}
We experiment with (post-deformation) SNRs of $-10$ dB, 0 dB, and 10 dB. 
The observed projections with the volume side length $N = 64$ are displayed in Fig. \ref{fig:projections_noise}.
The deformations comprise shifts between $\pm10$ pixels, shears between $\pm10\%$ of the sensor array, and rotation between $\pm10$ degrees.
We run $1500$ iterations of Adam to solve \eqref{eq:optim} with $\lambda_1=10$, $\lambda_2=1$, $\lambda_\theta=10^{-5}$ and $\lambda_\xb=10^{-5}$. 

\subsubsection{Results}

\begin{table}[b!]
	\centering
	\caption{Average error on deformation of the volume in Fig. \ref{fig:molecule_volume}. \label{tab:deformation_noise}}
	\begin{tabular}[t]{@{}rccc@{}}
		\toprule[1.3pt]
		&  \textbf{shift} [px] & \textbf{shear} [\%] & \textbf{rotation} [deg] \\
		\midrule
		\textit{Init} & 3.36 & 5.1 & 5.3  \\
		${-10}$ {dB}  & 0.86 & 5.8 & 4.0 \\
		${0}$ {dB} & 0.56 & 3.7 & 2.2 \\ 
		${10}$ {dB} & 0.36 & 2.9 & 1.4 \\ 
		\bottomrule[1.3pt]
	\end{tabular}
\end{table}%

\def\wid{2.5cm}
\def\hs{-1cm}
\def\xx{4.3}
\def\yy{0}
\def\zoomy{1.5}
\def\texty{-0.4}
\def\lw{1pt}
\def\hei{0.24}
\def\widfsc{0.28}
\def\sc{0.09}
\begin{figure*}[h!]
\begin{subfigure}{\linewidth}
	\hspace{0cm}
	\centering
	\begin{tikzpicture}[spy using outlines={circle,yellow,magnification=2,size=2.5cm, connect spies}]
	
	\node[rotate=270] at (0*\xx,0) { \includegraphics[width=\wid]{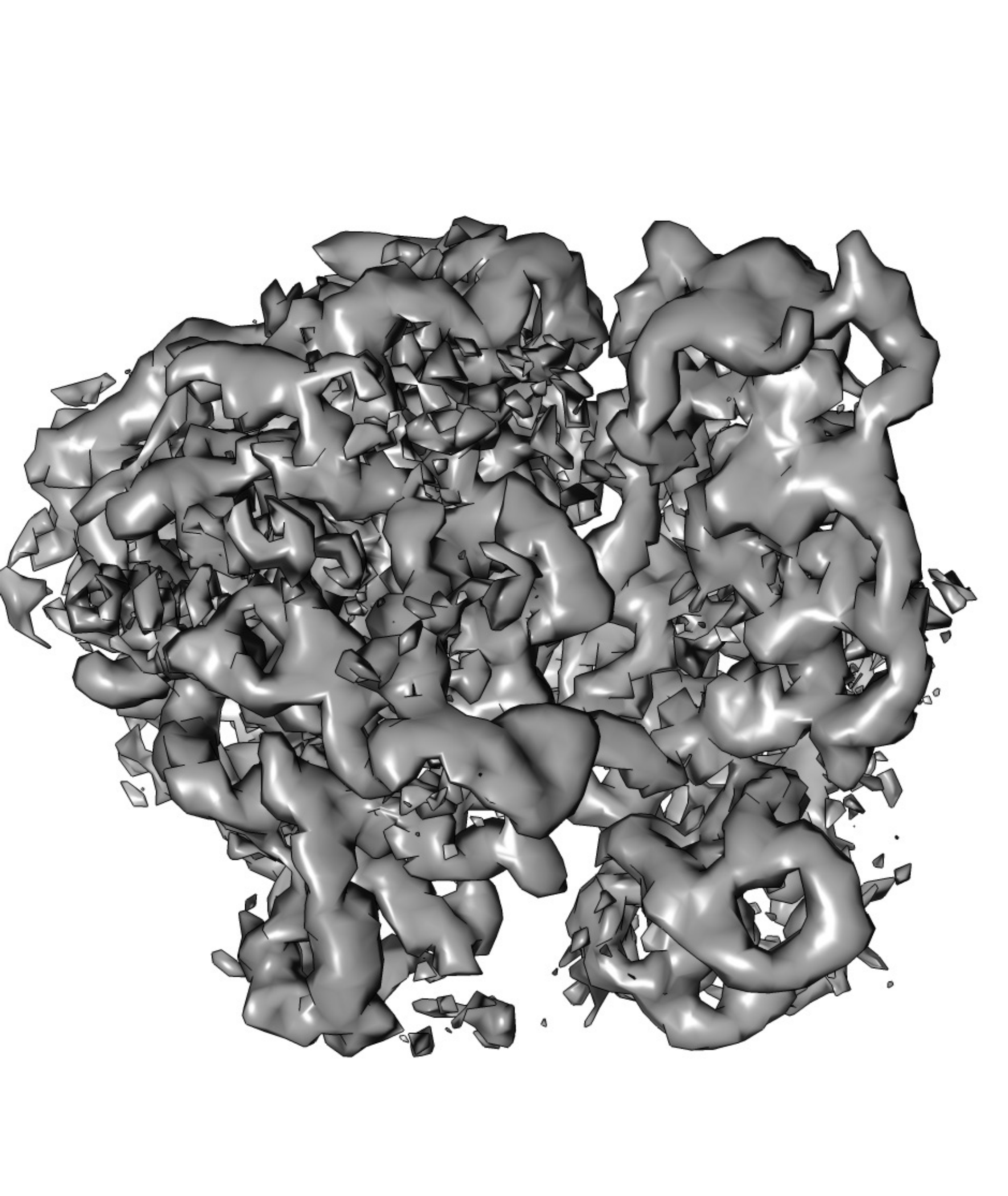}  };
	\node[rotate=270] at (1*\xx,0) { \includegraphics[width=\wid]{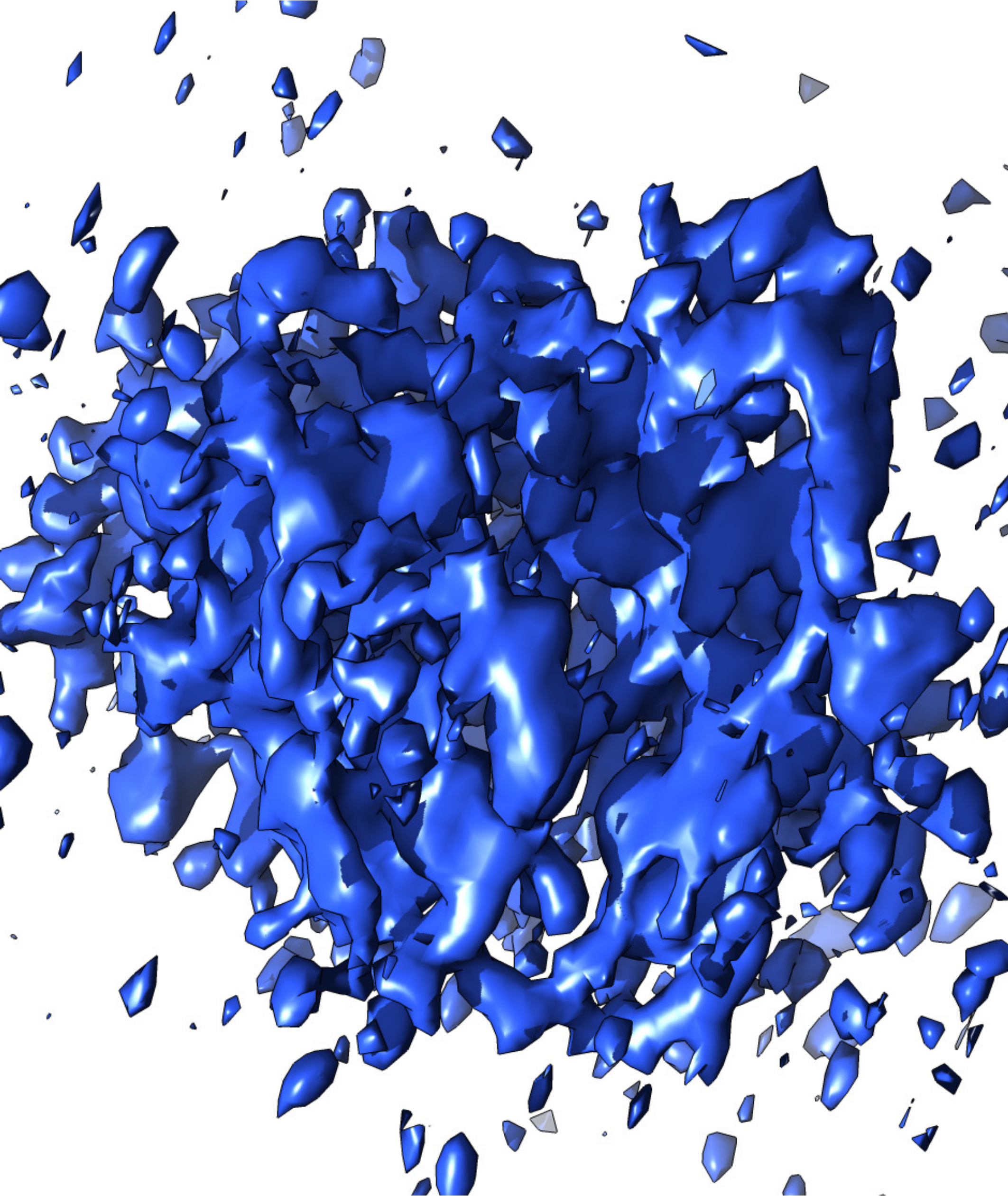}  };
	\node[rotate=270] at (2*\xx,0+\yy) { \includegraphics[width=\wid]{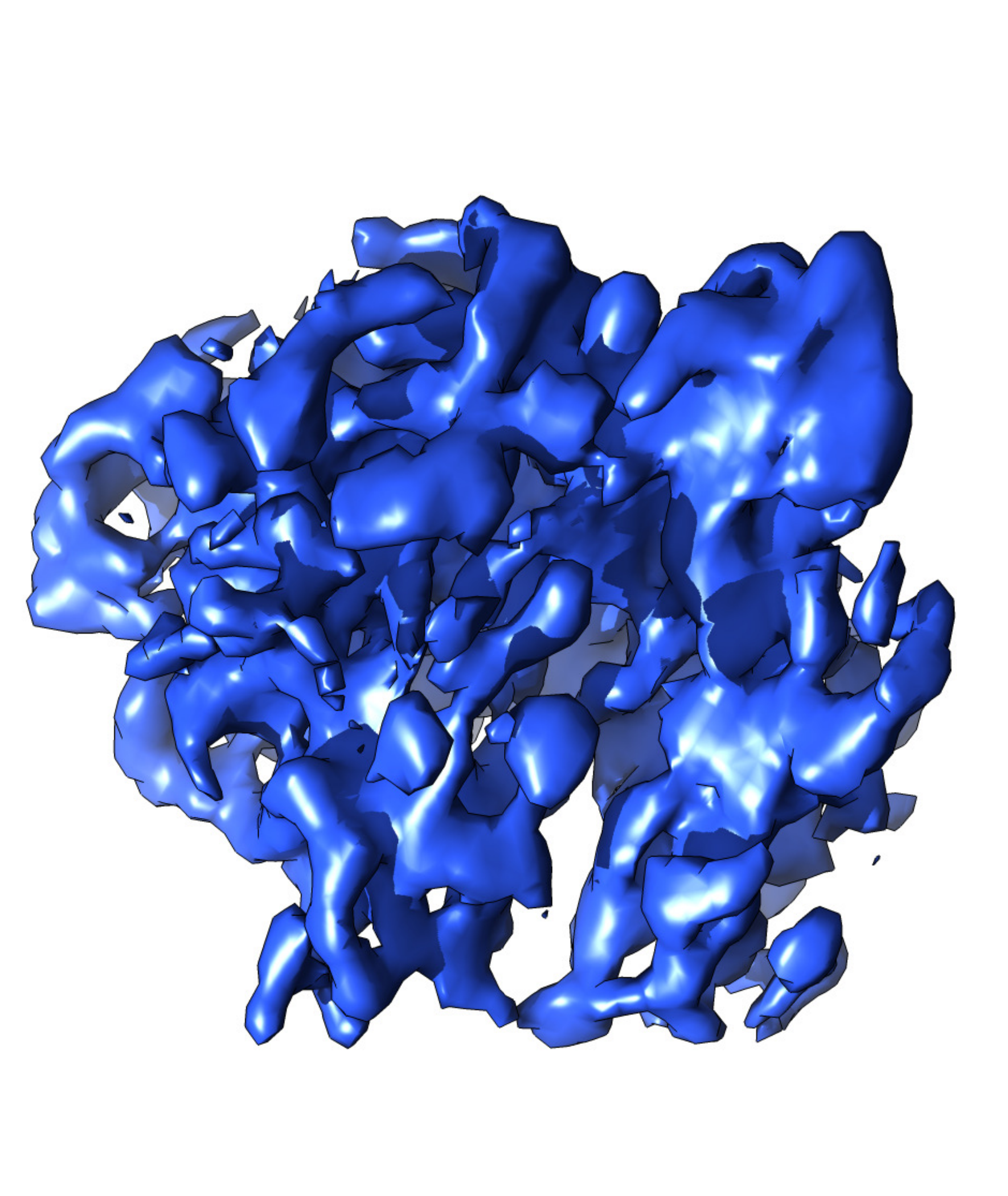}  };
	\node[rotate=270] at (3*\xx,0+\yy) { \includegraphics[width=\wid]{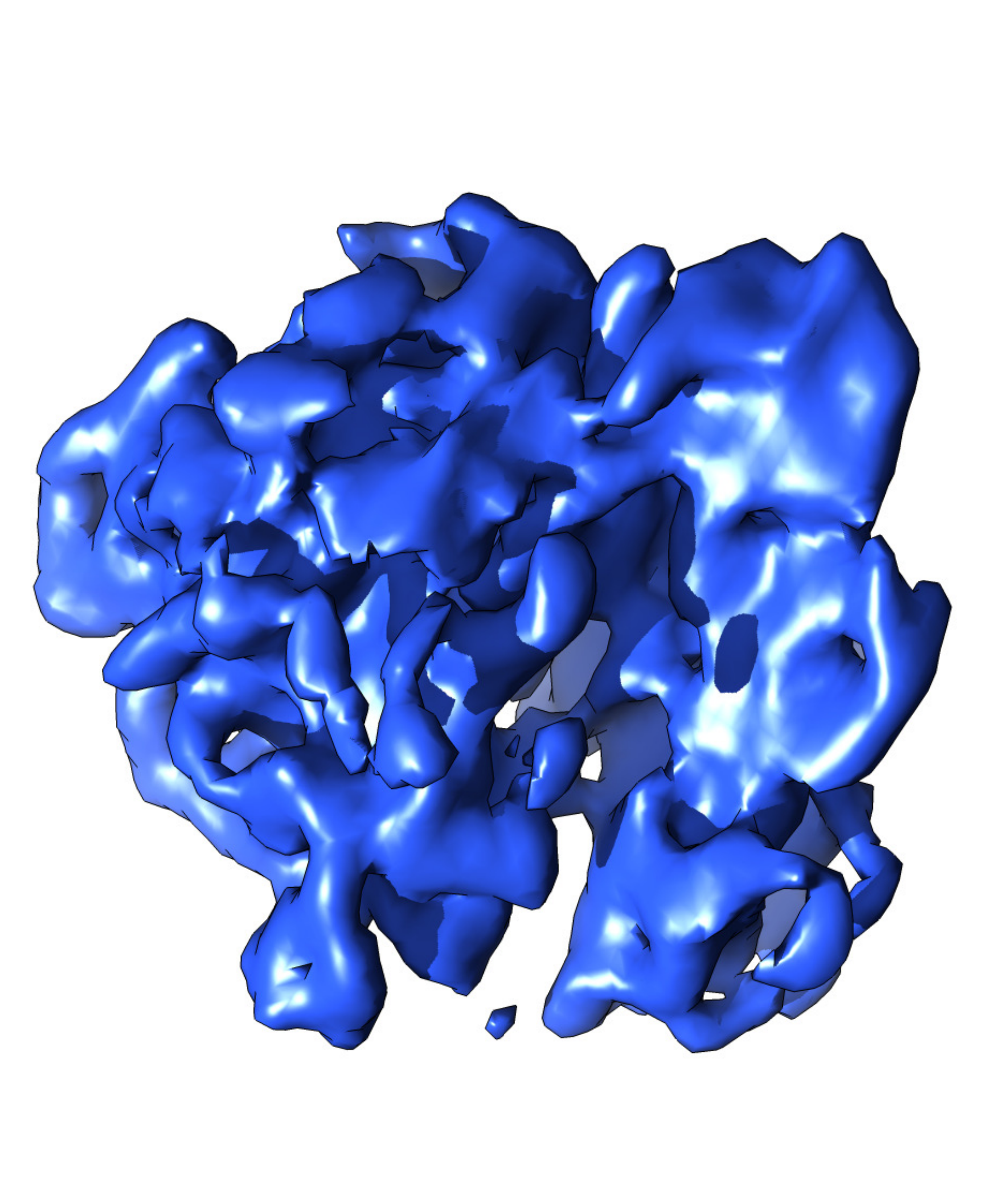}  };
	
	\spy on (-0.6,-0.3) in node [left] at (0,\zoomy);
	\spy on (-0.6+1*\xx,-0.3) in node [left] at (0+1*\xx,\zoomy);
	\spy on (-0.6+2*\xx,-0.3+\yy) in node [left] at (0+2*\xx,\zoomy+\yy);
	\spy on (-0.6+3*\xx,-0.3+\yy) in node [left] at (0+3*\xx,\zoomy+\yy);
	
	\node[text width=2cm,text centered,minimum width=2cm,minimum height=2cm] at (0.*\xx,3.5+\texty) {\textbf{True}};
	\node[text width=2cm,text centered,minimum width=2cm,minimum height=2cm] at (1*\xx,3.5+\texty) {\textbf{SNR -10 dB}};
	\node[text width=2cm,text centered,minimum width=2cm,minimum height=2cm] at (2*\xx,3.5+\yy+\texty) {\textbf{SNR 0 dB}};
	\node[text width=2cm,text centered,minimum width=2cm,minimum height=2cm] at (3*\xx,3.5+\yy+\texty) {\textbf{SNR 10 dB}};
	\end{tikzpicture} 
	\caption{\label{fig:molecule_volume}} 
	\vspace{-0.5cm}
\end{subfigure}
\begin{subfigure}{1\linewidth}
	\centering
	\begin{tikzpicture}  
		\node[text width=2cm,text centered,minimum width=2cm,minimum height=2cm] at (1.7,3.3) {SNR -10 dB};   
	\begin{axis}[
	width=\widfsc\linewidth, 
	height=\hei*\linewidth,
	grid=major, 
	grid style={dashed,gray!30}, 
	xlabel= frequency,
	ylabel=FSC,
	xmin = 0,
	ymin = 0,
	ytick={0,0.25,0.5,0.75,1},
	axis x line*=bottom,
	axis y line*=left,
	legend style={at={(1.0,1)}, legend cell align=left, align=left, draw=none,font=\scriptsize},
	yticklabel style={
		/pgf/number format/fixed,
		/pgf/number format/precision=1,
		/pgf/number format/fixed zerofill
	},
	scaled y ticks=false
	]
	\addplot[dashed, color=mycolor1,line width=\lw] table [col sep=comma] {images/SNR-10_scale2_proj60/fsc.csv};
	\addlegendentry{Estimated}
	\addplot[dashdotted, color=mycolor2,line width=\lw] table [col sep=comma] {images/SNR-10_scale2_proj60/fsc_fbp.csv};
	\addlegendentry{FBP}
	\addplot[dotted,color=mycolor3,line width=\lw] table [col sep=comma] {images/SNR-10_scale2_proj60/fsc_fbp_no_deformation.csv};
	\addlegendentry{FBP no deformation}
	\end{axis}
	\end{tikzpicture}
	\hfill
	\begin{tikzpicture}  
		\node[text width=2cm,text centered,minimum width=2cm,minimum height=2cm] at (1.7,3.3) {SNR 0 dB};   
	\begin{axis}[
	width=\widfsc\linewidth, 
	height=\hei*\linewidth,
	grid=major, 
	grid style={dashed,gray!30}, 
	xlabel= frequency,
	ylabel=FSC,
	xmin = 0,
	ymin = 0,
	ytick={},
	ytick={0,0.25,0.5,0.75,1},
	axis x line*=bottom,
	axis y line*=left,
	legend style={at={(1.1,0.95)}, legend cell align=left, align=left, draw=none,font=\scriptsize},
	yticklabel style={
		/pgf/number format/fixed,
		/pgf/number format/precision=1,
		/pgf/number format/fixed zerofill
	},
	scaled y ticks=false
	]
	\addplot[dashed,color=mycolor1,line width=\lw] table [col sep=comma] {images/SNR0_scale2_proj60/fsc.csv};
	\addplot[dashdotted,color=mycolor2,line width=\lw] table [col sep=comma] {images/SNR0_scale2_proj60/fsc_fbp.csv};
	\addplot[dotted,color=mycolor3,line width=\lw] table [col sep=comma] {images/SNR0_scale2_proj60/fsc_fbp_no_deformation.csv};
	\end{axis}
	\end{tikzpicture}
	\hfill
	\begin{tikzpicture}  
		\node[text width=2cm,text centered,minimum width=2cm,minimum height=2cm] at (1.7,3.3) {SNR 10 dB};   
	\begin{axis}[
	width=\widfsc\linewidth, 
	height=\hei*\linewidth,
	grid=major, 
	grid style={dashed,gray!30}, 
	xlabel= frequency,
	ylabel=FSC,
	xmin = 0,
	ymin = 0,
	ytick={},
	ytick={0,0.25,0.5,0.75,1},
	axis x line*=bottom,
	axis y line*=left,
	legend style={at={(1.1,0.95)}, legend cell align=left, align=left, draw=none,font=\scriptsize},
	yticklabel style={
		/pgf/number format/fixed,
		/pgf/number format/precision=1,
		/pgf/number format/fixed zerofill
	},
	scaled y ticks=false
	]
	\addplot[dashed,color=mycolor1,line width=\lw] table [col sep=comma] {images/SNR10_scale2_proj60/fsc.csv};
	\addplot[dashdotted,color=mycolor2,line width=\lw] table [col sep=comma] {images/SNR10_scale2_proj60/fsc_fbp.csv};
	\addplot[dotted,color=mycolor3,line width=\lw] table [col sep=comma] {images/SNR10_scale2_proj60/fsc_fbp_no_deformation.csv};
	\end{axis}
	\end{tikzpicture}
	\caption{\label{fig:fsc_noise}}
\end{subfigure}
\begin{subfigure}{1\linewidth}
	\centering
	\def\xxx{1.7}
	\begin{tikzpicture}
	\node at (0*\xxx,0) { \includegraphics[width=\sc\linewidth]{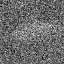} };
	\node at (1*\xxx,0) { \includegraphics[width=\sc\linewidth]{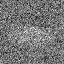} };
	\node at (2*\xxx,0) { \includegraphics[width=\sc\linewidth]{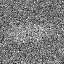} };
	
	\node at (3.2*\xxx,0) { \includegraphics[width=\sc\linewidth]{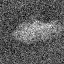} };
	\node at (4.2*\xxx,0) { \includegraphics[width=\sc\linewidth]{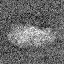} };
	\node at (5.2*\xxx,0) { \includegraphics[width=\sc\linewidth]{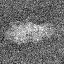} };
	
	\node at (6.4*\xxx,0) { \includegraphics[width=\sc\linewidth]{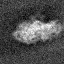} };
	\node at (7.4*\xxx,0) { \includegraphics[width=\sc\linewidth]{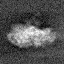} };
	\node at (8.4*\xxx,0) { \includegraphics[width=\sc\linewidth]{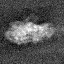} };
	
	\node[text = white, text centered] at (-0.1*\xxx,-0.4*\xxx) {{\scriptsize\textbf{SNR -10 dB}}};
	\node[text = white, text centered] at (3.1*\xxx,-0.4*\xxx) {{\scriptsize\textbf{SNR 0 dB}}};
	\node[text = white, text centered] at (6.3*\xxx,-0.4*\xxx) {{\scriptsize\textbf{SNR 10 dB}}};
	\end{tikzpicture} 
	\caption{\label{fig:projections_noise}}
	\vspace{-0.2cm}
\end{subfigure}
\caption{a) 3D density estimation at different SNR. b) FSC of the proposed approach compared to FBP reconstruction when measurements are perturbed or not by deformations. c) Three projections corresponding to consecutive viewing directions.}
\end{figure*}
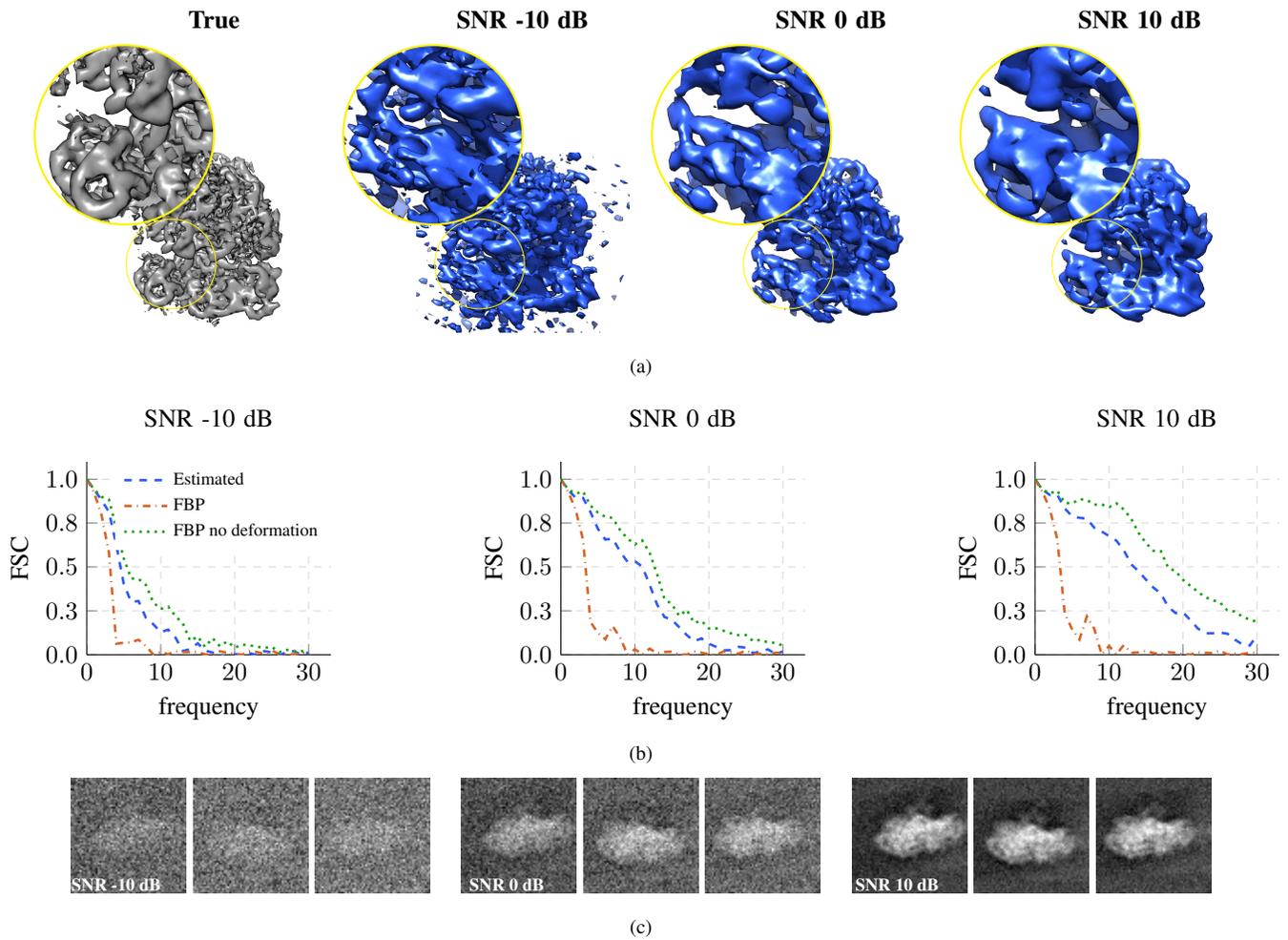

In Fig. \ref{fig:fsc_noise}, we report the Fourier Shell Correlation (FSC), a common metric to assess the quality of CryoET reconstruction \cite{harauz1986exact}. The FSC measures the correlation between the frequencies of the estimated volume and the original volume. We see that our method provides a significant gain compared to directly applying the FBP reconstruction algorithm on the raw observation. Even with large measurement noise we successfully recover projections close to the non-deformed ones. 

In Table \ref{tab:deformation_noise} we report the average error over the $M$ projections between the true and the estimated deformation parameters given by solving \eqref{eq:optim}. 
The initialization for the deformation parameters is $\phib = 0$, that is to say, no deformation. 
This quantitative inspection confirms the ability of the proposed approach to identify the deformation parameters. 
We display the reconstructed volume in Fig. \ref{fig:molecule_volume}. While the overall structure is well-retrieved at reasonable SNRs, we observe a severe loss of fine details at SNR -10 dB. However, at SNR 0 dB and 10 dB, as indicated by the FSC scores, the smallest details are correctly reconstructed; see insets in Fig. \ref{fig:molecule_volume}.

\subsection{Volume density from a real CryoET acquisition}

Finally, we experiment on a volume obtained using in a real CryoET acquisition. We use the cryo-electron tomogram of mouse hippocampal neurons \cite{nedozralova2022situ} (dataset DOI on EMPIAR 10.6019/EMPIAR-10923) display in Fig. \ref{fig:neuron_volume}.

\subsubsection{Experimental parameters}
We simulate CryoET acquisition \eqref{eq:CryoET_formulation} by collecting $50$ projections from angles between $-70$ to $70$ degrees. 
The measurement SNR, after adding noise to the deformed projections, is 10 dB.
Note that this SNR is comparatively more severe for this experiment than in the previous experiments because the volume is much less sparse.
The volume density is of size $128\times 128 \times 90$ and the projections are of size $128\times 128$.
The deformations comprise shifts between $\pm5$ pixels, shear between $\pm5\%$ and rotation between $\pm5$ degrees.
We run $2000$ iterations of Adam to solve the optimization problem \eqref{eq:optim} with $\lambda_1=100$, $\lambda_2=10^{-2}$, $\lambda_\theta=10^{-6}$ and $\lambda_\xb=10^{-5}$.

\subsubsection{Results}
\begin{table}[b!]
\centering
\caption{Average error on deformation of the volume in Fig. \ref{fig:neuron_volume}. \label{tab:deformation_neuron} } 
\begin{tabular}[t]{@{}lccc@{}}
	\toprule[1.3pt]
	&  \textbf{shift} [px] & \textbf{shear} [\%] & \textbf{rotation} [deg] \\
	\midrule
	\textit{Initialization} & 1.7 & 2.5 & 2.7  \\
	\textit{Estimate}  & 0.2 & 1.9 & 1.1\\
	\bottomrule[1.3pt]
\end{tabular}
\end{table}%

We display several slices of the true and the estimated volume in Fig \ref{fig:neuron_volume}. The FSC in Fig. \ref{fig:neuron_fsc} shows that both the coarse structure and fine details are reconstructed well. We obtain accurate estimates of the deformation as shown in Table \ref{tab:deformation_neuron}.

\def\lw{1pt}
\def\xx{1.8}
\def\yy{-1.8}
\def\hei{1}
\def\sc{0.2}

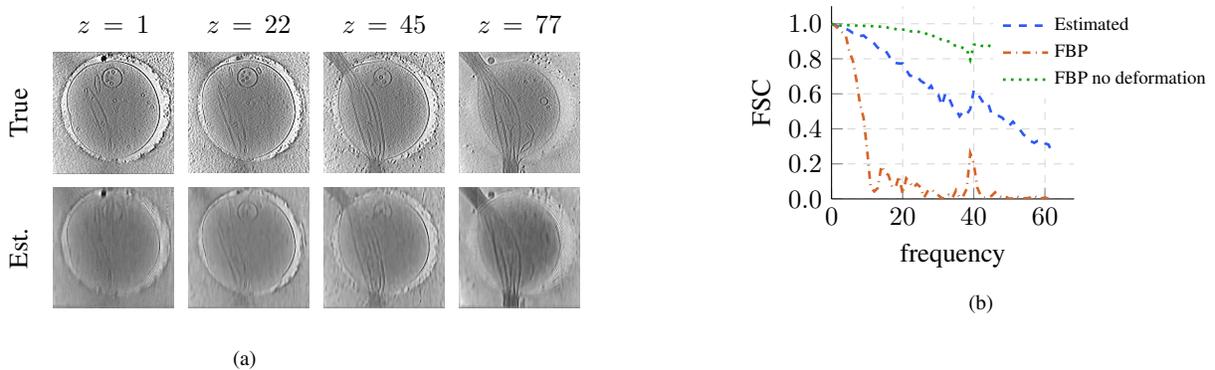
\begin{figure*}[h!]

\begin{subfigure}{0.45\linewidth}
\centering
\begin{tikzpicture} 
\node[rotate=0] at (0,0) {
	\includegraphics[width=\sc\linewidth]{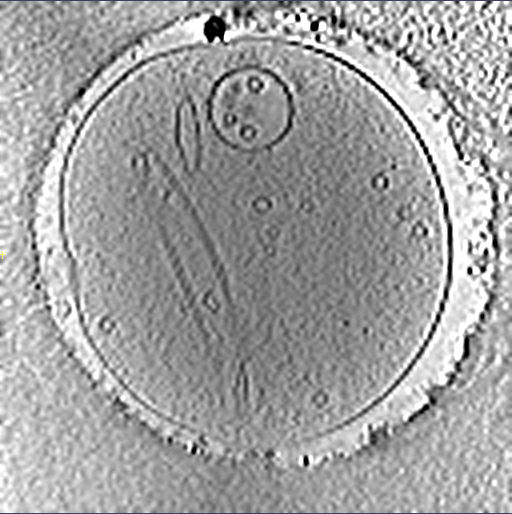}};
\node[rotate=0] at (\xx,0) {
	\includegraphics[width=\sc\linewidth]{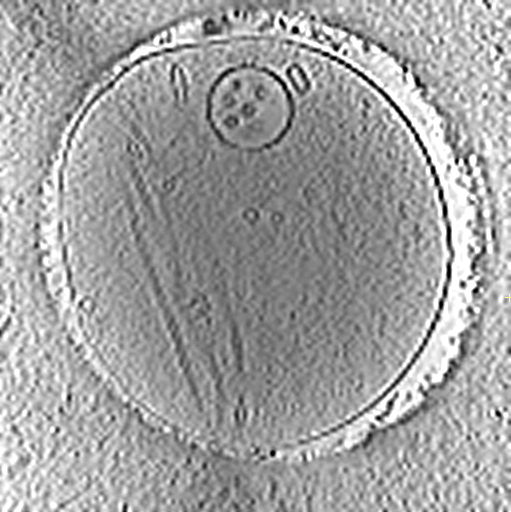}};
\node[rotate=0] at (2*\xx,0) {
	\includegraphics[width=\sc\linewidth]{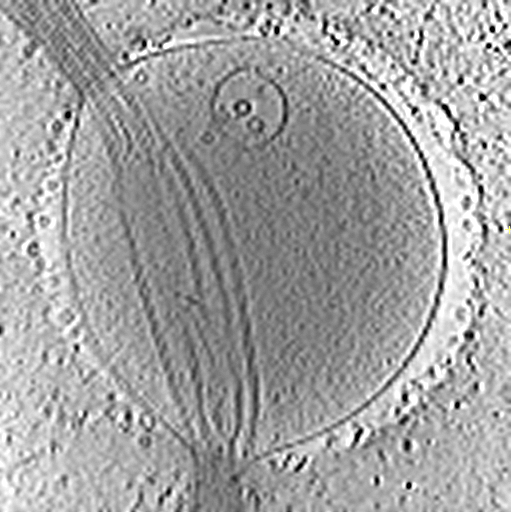}};
\node[rotate=0] at (3*\xx,0) {
	\includegraphics[width=\sc\linewidth]{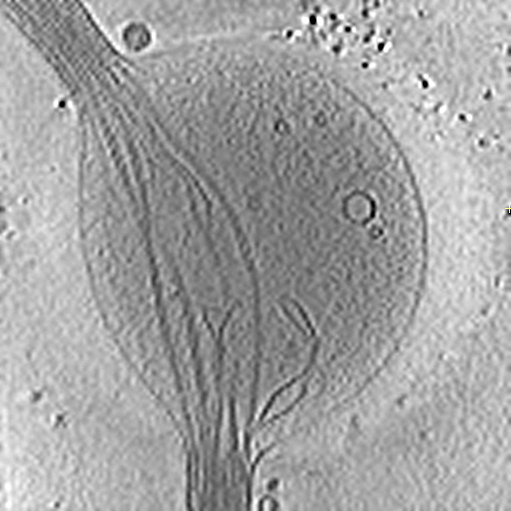}};
\vspace{0.1cm}
\node[yscale=-1,inner sep=0,outer sep=0,rotate=0] at (0,\yy) {
	\includegraphics[width=\sc\linewidth]{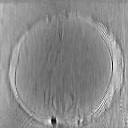}};
\node[yscale=-1,inner sep=0,outer sep=0,rotate=0] at (\xx,\yy) {
	\includegraphics[width=\sc\linewidth]{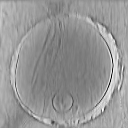}};
\node[yscale=-1,inner sep=0,outer sep=0,rotate=0] at (2*\xx,\yy) {
	\includegraphics[width=\sc\linewidth]{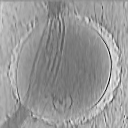}};
\node[yscale=-1,inner sep=0,outer sep=0,rotate=0] at (3*\xx,\yy) {
	\includegraphics[width=\sc\linewidth]{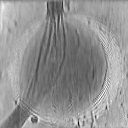}};

\node[text width=2cm,text centered,minimum width=2cm,minimum height=2cm, rotate=90] at 	(-0.7*\xx,0) {True};
\node[text width=2cm,text centered,minimum width=2cm,minimum height=2cm, rotate=90] at 	(-0.7*\xx,\yy) {Est.};
\node[text width=2cm,text centered,minimum width=2cm,minimum height=2cm, rotate=0] at (0,0.65*\xx) {$z=1$};
\node[text width=2cm,text centered,minimum width=2cm,minimum height=2cm, rotate=0] at (\xx,0.65*\xx) {$z=22$};
\node[text width=2cm,text centered,minimum width=2cm,minimum height=2cm, rotate=0] at (2*\xx,0.65*\xx) {$z=45$};
\node[text width=2cm,text centered,minimum width=2cm,minimum height=2cm, rotate=0] at (3*\xx,0.65*\xx) {$z=77$};
\end{tikzpicture} 
\caption{
\label{fig:neuron_volume}}
\end{subfigure}\hfill
\begin{subfigure}{0.45\linewidth}
\centering
\begin{tikzpicture}  
\begin{axis}[
width=0.6\linewidth, 
grid=major, 
grid style={dashed,gray!30}, 
xlabel= frequency,
ylabel=FSC,
xmin = 0,
ymin = -0.00,
ytick={0,0.2,0.4,0.6,0.8,1},
axis x line*=bottom,
axis y line*=left,
legend style={at={(1.6,1.0)}, legend cell align=left, align=left, draw=none,font=\scriptsize},
yticklabel style={
	/pgf/number format/fixed,
	/pgf/number format/precision=1,
	/pgf/number format/fixed zerofill
},
scaled y ticks=false
]
\addplot[dashed, color=mycolor1,line width=\lw] table [col sep=comma] {images/neuron/fsc.csv};
\addlegendentry{Estimated}
\addplot[dashdotted, color=mycolor2,line width=\lw] table [col sep=comma] {images/neuron/fsc_fbp.csv};
\addlegendentry{FBP}
\addplot[dotted,color=mycolor3,line width=\lw] table [col sep=comma] {images/neuron/fsc_fbp_no_deformation.csv};
\addlegendentry{FBP no deformation}
\end{axis}
\end{tikzpicture}
\caption{
\label{fig:neuron_fsc}}
\end{subfigure}
\caption{a) Reconstruction of CryoET volume~\cite{nedozralova2022situ} (second row) at different depths ($z$), compared with the original volume (first row). b) FSC for the neuron volume of Fig. \ref{fig:neuron_volume} for the proposed approach compared to FBP reconstruction when measurements are perturbed or not by deformations.}
\end{figure*}

\section{Conclusions}

We demonstrated how coordinate-based neural representations can be made into an effective tool for joint projection alignment and calibration in CryoET. The fact that we can seamlessly build diverse deformation models as coordinate transformations in neural fields and then automatically differentiate with respect to their parameters gives us a simple and powerful framework. Ongoing work includes extensions of deformation classes to model the full complexity of those encountered in CryoET (although the three we use are known to be the most important ones). In order to keep complexity under control it may be helpful to borrow the parametric classes used to model spatially-varying blur in light microscopy \cite{debarnot2019scalable, blindCNN}. Here we again benefit from the flexibility of the introduced framework.


\vfill\pagebreak
\newpage
\FloatBarrier

\bibliographystyle{IEEEbib}
\bibliography{refs}

\begin{thebibliography}{10}

\bibitem{mastronarde2007fiducial}
David~N Mastronarde,
\newblock ``Fiducial marker and hybrid alignment methods for single-and
  double-axis tomography,''
\newblock in {\em Electron tomography}, pp. 163--185. Springer, 2007.

\bibitem{harauz1986exact}
George Harauz and Marin van Heel,
\newblock ``Exact filters for general geometry three dimensional
  reconstruction.,''
\newblock {\em Optik.}, vol. 73, no. 4, pp. 146--156, 1986.

\bibitem{mastronarde2017automated}
David~N Mastronarde and Susannah~R Held,
\newblock ``Automated tilt series alignment and tomographic reconstruction in
  imod,''
\newblock {\em Journal of structural biology}, vol. 197, no. 2, pp. 102--113,
  2017.

\bibitem{fernandez2021tomoalign}
Jose-Jesus Fernandez and Sam Li,
\newblock ``Tomoalign: A novel approach to correcting sample motion and 3d ctf
  in cryoet,''
\newblock {\em Journal of Structural Biology}, vol. 213, no. 4, pp. 107778,
  2021.

\bibitem{tegunov2019real}
Dimitry Tegunov and Patrick Cramer,
\newblock ``Real-time cryo-electron microscopy data preprocessing with warp,''
\newblock {\em Nature methods}, vol. 16, no. 11, pp. 1146--1152, 2019.

\bibitem{gupta2022differentiable}
Sidharth Gupta, Konik Kothari, Valentin Debarnot, and Ivan Dokmani{\'c},
\newblock ``Differentiable uncalibrated imaging,''
\newblock {\em arXiv preprint arXiv:2211.10525}, 2022.

\bibitem{xie2022neural}
Yiheng Xie, Towaki Takikawa, Shunsuke Saito, Or~Litany, Shiqin Yan, Numair
  Khan, Federico Tombari, James Tompkin, Vincent Sitzmann, and Srinath Sridhar,
\newblock ``Neural fields in visual computing and beyond,''
\newblock in {\em Computer Graphics Forum}. Wiley Online Library, 2022,
  vol.~41, pp. 641--676.

\bibitem{zheng2022aretomo}
Shawn Zheng, Georg Wolff, Garrett Greenan, Zhen Chen, Frank~GA Faas, Montserrat
  B{\'a}rcena, Abraham~J Koster, Yifan Cheng, and David~A Agard,
\newblock ``Aretomo: An integrated software package for automated marker-free,
  motion-corrected cryo-electron tomographic alignment and reconstruction,''
\newblock {\em Journal of Structural Biology: X}, vol. 6, pp. 100068, 2022.

\bibitem{liu2022mechanical}
Yan Liu, Jonathan Dong, Thanh-an Pham, Fran{\'{c}}ois Marelli, and Michael
  Unser,
\newblock ``Mechanical artifacts in optical projection tomography:
  Classification and automatic calibration,''
\newblock {\em arXiv preprint arXiv:2210.03513}, 2022.

\bibitem{mildenhall2020nerf}
Ben Mildenhall, Pratul~P Srinivasan, Matthew Tancik, Jonathan~T Barron, Ravi
  Ramamoorthi, and Ren Ng,
\newblock ``Nerf: Representing scenes as neural radiance fields for view
  synthesis,''
\newblock in {\em European conference on computer vision}. Springer, 2020, pp.
  405--421.

\bibitem{sitzmann2020implicit}
Vincent Sitzmann, Julien Martel, Alexander Bergman, David Lindell, and Gordon
  Wetzstein,
\newblock ``Implicit neural representations with periodic activation
  functions,''
\newblock {\em Advances in Neural Information Processing Systems}, vol. 33, pp.
  7462--7473, 2020.

\bibitem{sun2021coil}
Yu~Sun, Jiaming Liu, Mingyang Xie, Brendt Wohlberg, and Ulugbek~S Kamilov,
\newblock ``Coil: Coordinate-based internal learning for tomographic imaging,''
\newblock {\em IEEE Transactions on Computational Imaging}, vol. 7, pp.
  1400--1412, 2021.

\bibitem{tegunov2021multi}
Dimitry Tegunov, Liang Xue, Christian Dienemann, Patrick Cramer, and Julia
  Mahamid,
\newblock ``Multi-particle cryo-em refinement with m visualizes
  ribosome-antibiotic complex at 3.5 {\aa} in cells,''
\newblock {\em Nature Methods}, vol. 18, no. 2, pp. 186--193, 2021.

\bibitem{nedozralova2022situ}
Hana Nedozralova, Nirakar Basnet, Iosune Ibiricu, Satish Bodakuntla, Christian
  Biert{\"u}mpfel, and Naoko Mizuno,
\newblock ``In situ cryo-electron tomography reveals local cellular machineries
  for axon branch development,''
\newblock {\em Journal of Cell Biology}, vol. 221, no. 4, 2022.

\bibitem{debarnot2019scalable}
Valentin Debarnot, Paul Escande, and Pierre Weiss,
\newblock ``A scalable estimator of sets of integral operators,''
\newblock {\em Inverse Problems}, vol. 35, no. 10, pp. 105011, 2019.

\bibitem{blindCNN}
Valentin Debarnot and Pierre Weiss,
\newblock ``{Deep-Blur : Blind Identification and Deblurring with Convolutional
  Neural Networks},''
\newblock preprint, 2022.

\end{thebibliography}

\end{document}